\documentclass[]{emulateapj}

\let\lsim=\la

\slugcomment{Accepted for publication in the Astrophysical Journal Letters}
\shorttitle{ALMA 1.3~mm Number Counts}
\shortauthors{Hatsukade et al.}

\begin{document}

\title{Faint End of 1.3~mm Number Counts Revealed by ALMA}

\author{
	Bunyo Hatsukade\altaffilmark{1},
	Kouji Ohta\altaffilmark{1},
	Akifumi Seko\altaffilmark{1},
	Kiyoto Yabe\altaffilmark{2},
	and 
	Masayuki Akiyama\altaffilmark{3},
}

\affil{\altaffilmark{1} Department of Astronomy, Kyoto University, Kyoto 606-8502, Japan}
\affil{\altaffilmark{2} National Astronomical Observatory of Japan, 2-21-1 Osawa, Mitaka, Tokyo 181-8588, Japan}
\affil{\altaffilmark{3} Astronomical Institute, Tohoku University, Aramaki, Aoba-ku, Sendai, Miyagi 980-8578, Japan}
\altaffiltext{1}{hatsukade@kusastro.kyoto-u.ac.jp}

\begin{abstract}
We present the faint end of number counts at 1.3~mm (238~GHz) obtained with the Atacama Large Millimeter/submillimeter Array (ALMA). 
Band 6 observations were carried out targeting 20 star-forming galaxies at $z \sim 1.4$ in the Subaru/{\sl XMM-Newton} Deep Survey field. 
In the observations, we serendipitously detect 15 sources ($\ge$$3.8 \sigma$, $S_{\rm 1.3\ mm} = 0.15$--0.61 mJy) other than the targeted sources. 
We create number counts by using these `sub-mJy sources', which probe the faintest flux range among surveys at millimeter wavelengths. 
The number counts are consistent with (flux-scaled) number counts at 850~$\mu$m and 870~$\mu$m obtained with gravitational lensing clusters. 
The ALMA number counts agree well with model predictions, which suggest that these sub-mJy populations are more like `normal' star-forming galaxies than `classical' SMGs with intense star-forming activity. 
In this flux range, $\sim$80\% of the extragalactic background light at 1.3~mm is resolved into individual sources. 
\end{abstract}


\section{Introduction}

Extragalactic surveys at millimeter and submillimeter have uncovered a population of dusty star-forming galaxies at high redshifts, so-called submillimeter galaxies (SMGs). 
SMGs are dusty starburst galaxies with star-formation rates (SFRs) of $10^2$--$10^3$~$M_{\odot}$~yr$^{-1}$ \citep[e.g.,][]{blai02}. 
SMGs are important for understanding of the process of galaxy formation, the cosmic star-formation history, and the origin of extragalactic background light (EBL). 
The EBL is the integral of unresolved emission from extragalactic sources and contains information on the history of galaxy evolution. 
A large fraction of the EBL at millimeter and submillimeter wavelengths is thought to be contributed by high-redshift dusty galaxies \citep{laga05}. 
Previous blank field surveys resolved 20\%--40\% of the EBL at 850~$\mu$m \citep[e.g.,][]{eale00, copp06} and 10\%--20\% at 1~mm \citep[e.g.,][]{grev04, pere08, scot10, hats11}. 
Meanwhile, in cluster fields 50\%--100\% of the EBL are resolved by utilizing the effect of gravitational lensing \citep[e.g.,][]{smai97, cowi02, knud08}, suggesting that a large fraction of the EBL originates from faint sources ($S_{\rm 1\ mm} < 1$ mJy) of which number counts are not yet well constrained due to the limited sensitivity and confusion limit for the single-dish surveys. 
Compared to SMGs detected in previous surveys, fainter submillimeter sources are possibly connected to more general star-forming galaxies, or galaxies with hotter dust.
Observations of fainter sources are needed to understand the formation and evolution of galaxies.

The Atacama Large Millimeter/submillimeter Array (ALMA) enables us to explore much fainter sources without the effect of confusion limit because of its high sensitivity and high angular resolution. 
In this letter, we present the faint end of number counts at 1.3~mm revealed by ALMA band 6 observations. 
The arrangement of this paper is as follows. 
Section~\ref{sec:observations} outlines the observations and data reduction. 
In Section~\ref{sec:source}, we describe the method of source extraction and simulations carried out to create number counts. 
In Section~\ref{sec:discussion}, we compare the number counts with other observational results and model predictions, and estimate the contribution of the resolved 1.3~mm sources to the EBL. 
A summary is presented in Section~\ref{sec:summary}.

\section{Observations and Data Reduction}\label{sec:observations}

We conducted ALMA band 6 observations toward 20 star-forming galaxies at $z \sim 1.4$ in the Subaru XMM/Newton Deep Survey (SXDS) field \citep{furu08}. 
The targets were extracted from a stellar mass limit sample whose redshifts were obtained by near-infrared (NIR) spectroscopy with the Fibre Multi-Object Spectro-graph \citep[FMOS;][]{kimu10} on the Subaru telescope \citep[][]{yabe12, yabe13}. 
The observations were carried out in August 9th, 11th, 15th, and 26th, 2012 with 23--25 antennas in the extended configuration during the ALMA Cycle~0 session. 
The correlator was used in the frequency domain mode with a bandwidth of 1875 MHz (488.28 kHz $\times$ 3840 channels). 
The targets were divided into four groups with different spectral settings to observe the redshifted CO(5--4) line as efficiently as possible. 
The local oscillator frequencies for the four groups were 231.198~GHz, 236.168~GHz, 240.380~GHz, and 244.166~GHz (typical frequency of 238~GHz). 
Three or four spectral windows with a bandwidth of 1875 MHz, each with a different central frequency, were used in each group to cover the CO(5--4) line and simultaneously detect the dust thermal continuum emission. 
Ganymede, Callisto, and Uranus were observed as a flux calibrator. 
The bandpass and phase were calibrated with J2253$+$161 and J0204$-$170, respectively. 
We obtained 20 pointings centered on the 20 targets, each with on-source observing time of 8--15 minutes. 
The full width at half maximum (FWHM) of the primary beam is $26''$.

The data were reduced with the Common Astronomy Software Applications \citep[CASA;][]{mcmu07} package in a standard manner. 
The maps were processed with the {\verb CLEAN } algorithm with the natural weighting, which gives the final synthesized beamsize of $\sim$$0.6''$--$1.3''$.  
The rms noise level of the 20 maps is 0.04--0.10 mJy~beam$^{-1}$ and is almost constant in each map uncorrected for the primary beam attenuation. 
In what follows, we use the area within the primary beam of the 20 maps. 
In this letter, we focus on serendipitously detected sources other than the targeted sources to examine the source number counts. 
Details on the targeted star-forming galaxies are discussed elsewhere.

\section{Source Extraction and Number Counts}\label{sec:source}

\subsection{Source and Spurious Detection}\label{sec:spurious}

\begin{figure}
\vspace{5mm}
\includegraphics[width=\linewidth]{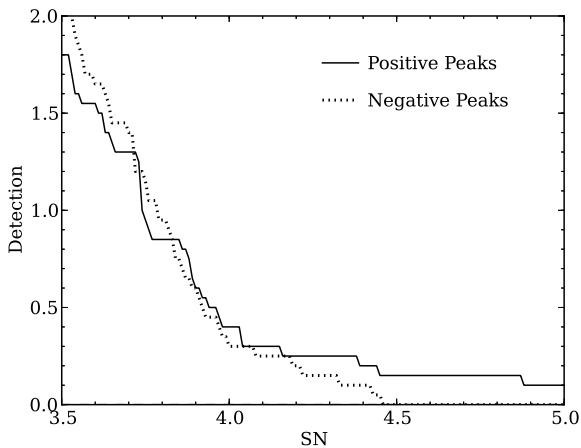}
\caption{
Average number of positive and negative peaks in a map as a function of SN threshold. 
\label{fig:fdr}}
\end{figure}

Source extraction is conducted on the 20 maps (before primary beam correction) where all the sources with a peak signal-to-noise ratio (SN) above 3.5 are {\verb CLEAN }ed. 
We adopt the detection threshold of $3.8 \sigma$, which is determined from the number of spurious sources expected in the maps. 
The number of spurious source is estimated by counting the negative peaks in each map.
Spurious source extraction is performed on the maps multiplied by $-1$. 
The spatial distribution of the spurious sources is almost uniform in each map. 
Fig.~\ref{fig:fdr} shows the number of positive and negative peaks in a map as a function of SN threshold averaged over the 20 fields. 
We find that the number of spurious sources is less than one at SN $\ge 3.8$.

\begin{figure}
\vspace{10mm}
\begin{center}
\includegraphics[width=.9\linewidth]{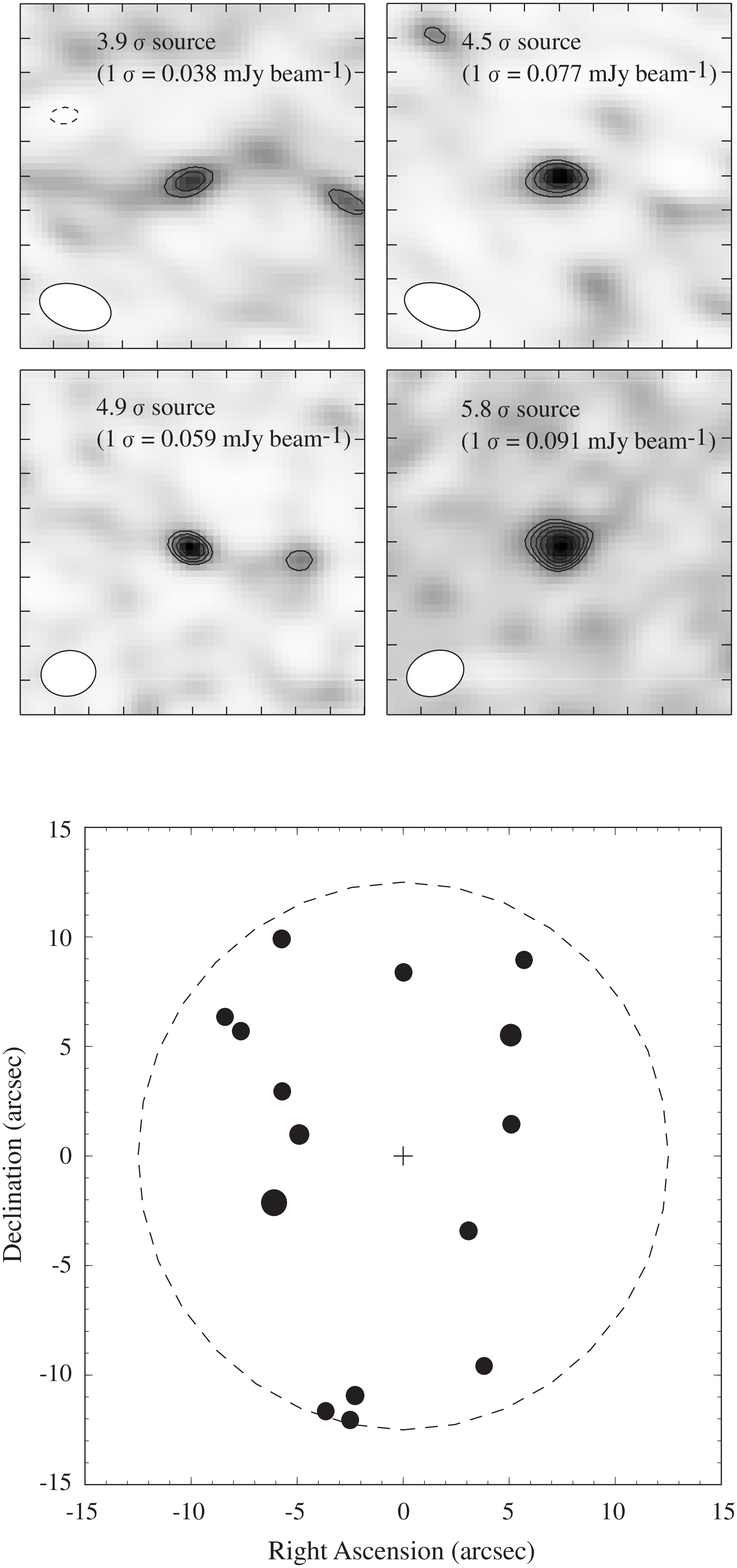}
\end{center}
\caption{
Top: Examples of detections in four maps ($5'' \times 5''$). 
Contours start from $\pm 3.0 \sigma$ with a $0.5 \sigma$ step. 
Negative contours are presented in dashed lines. 
The synthesized beam is shown in the bottom left in each map. 
Bottom: Positions of the 15 source in the maps. 
The size of each symbol is scaled by signal-to-noise ratios. 
Dashed circle represents the primary beam. 
\label{fig:source}}
\end{figure}

We detect 15 sources (excluding the target sources) in the 20 maps with the raw flux density (uncorrected for the primary beam attenuation) of 0.15--0.53 mJy (3.86--$5.84 \sigma$). 
All the sources are point sources or only marginally resolved. 
A few sources are possibly detected in emission lines which may dominate their flux density. 
Examples of detected sources in the maps with different SNs are shown in Fig.~\ref{fig:source} (top). 
Fig.~\ref{fig:source} (bottom) shows the source positions in the maps. 
No clear clustering feature is seen in the sky distribution of the sources.

\subsection{Completeness}\label{sec:completeness}

We calculate the completeness, which is the rate at which a source is expected to be detected in a map, to see the effect of noise fluctuation on the source detection. 
Because the source extraction is conducted in the maps uncorrected for primary beam attenuation, the completeness calculation for each source is also conducted in the same map by using the raw flux density. 
We subtract $\ge$$3.5 \sigma$ sources from each map, and inject a flux-scaled synthesized beam into the map. 
Because the rms noise level is almost constant in each map, the input position is randomly selected in the map. 
When the input source is extracted within $1.0''$\footnotemark of its input position with $\ge$$3.8 \sigma$, the source is considered to be recovered. 
We repeat this 1000 times for the raw flux density of each source. 
The completeness for the 15 sources ranges from 77\% to 100\%.

In the course of this simulation, we calculate the ratio between the input and output fluxes to estimate the intrinsic flux density of the detected sources. 
The flux ratio calculated for each source in this simulation is 1.00--1.14. 
The flux density of the sources corrected for this effect and the the primary beam attenuation is 0.15--0.61 mJy.

\footnotetext{Note that the effect of the choice of $0.5''$ or $2''$ on the number counts is within the errors. }

\subsection{Effective Area}\label{sec:area}

Because the rms noise levels are different in the 20 maps 
and the primary beam response in a map changes depending on the distance from the map center, the survey area for each source should be calculated depending on its intrinsic flux density (corrected for primary beam attenuation). 
We calculate the effective area against the intrinsic flux density in each map; 
By using the rms noise levels corrected for primary beam attenuation, we find a distance from the map center within which the source is detected at $\ge$$3.8 \sigma$ in the raw map uncorrected for primary beam attenuation.

\begin{figure}
\includegraphics[width=\linewidth]{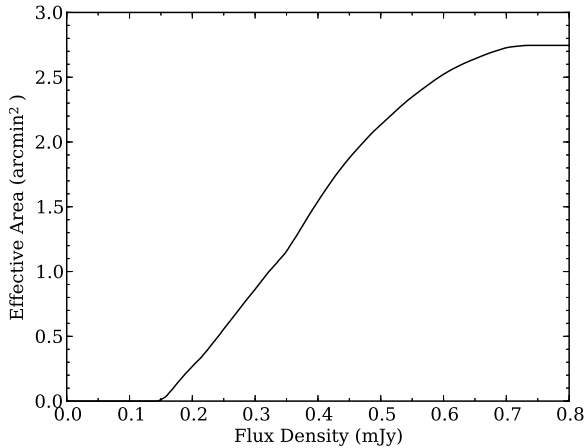}
\caption{
Effective area as a function of flux density, where a source with a flux density (corrected for primary beam attenuation) 
can be detected with $\ge$$3.8 \sigma$. 
\label{fig:area}}
\vspace{1mm}
\end{figure}

\subsection{Number Counts}\label{sec:counts}

\begin{figure}
\begin{center}
\includegraphics[width=\linewidth]{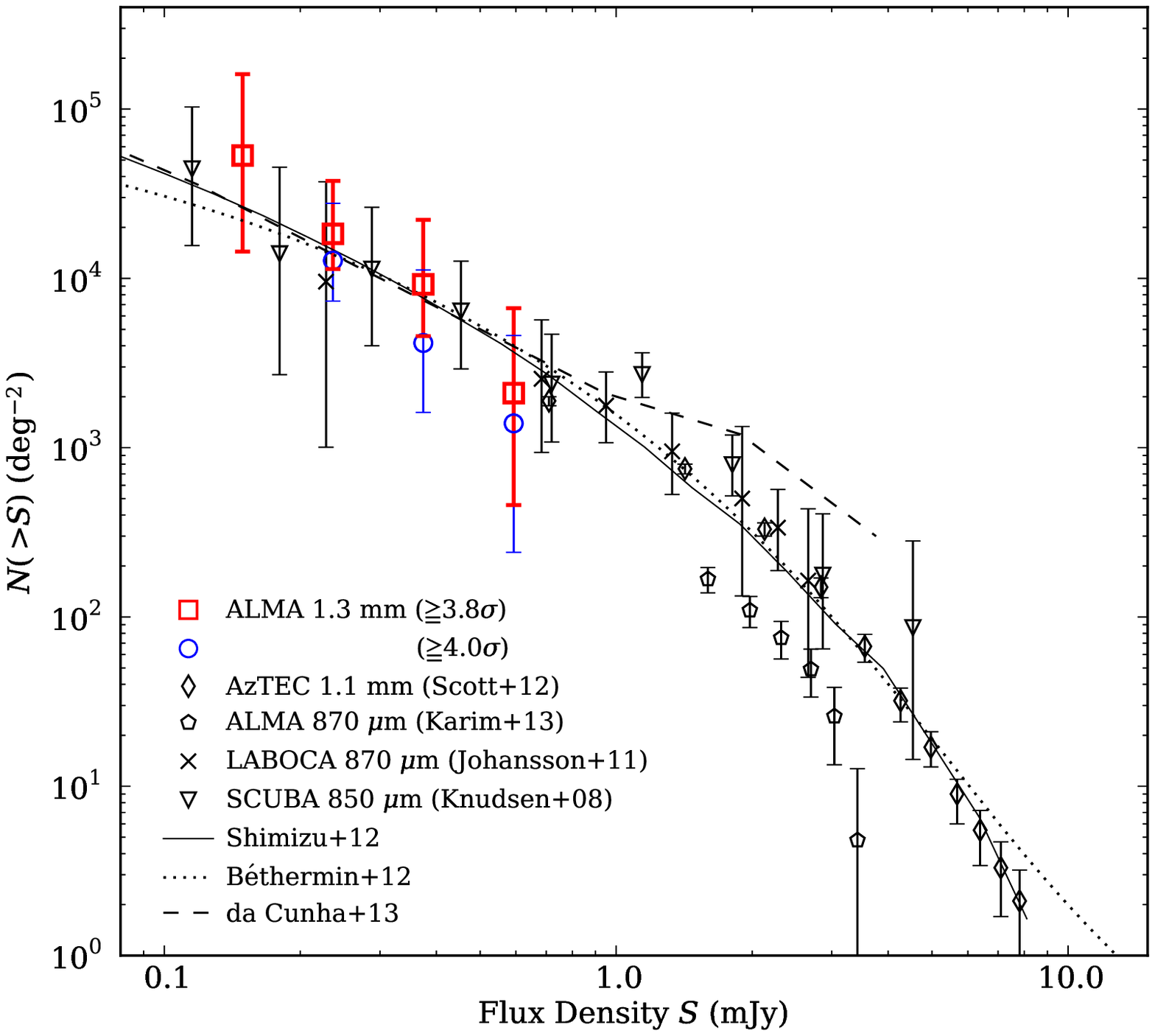}
\end{center}
\vspace{-4mm}
\caption{
Cumulative number counts at 1.3~mm obtained in the ALMA band 6 observations (red squares for $\ge$$3.8\sigma$ sources and blue circles for $\ge$$4.0\sigma$ sources). 
For comparison, we plot the ALMA 870~$\mu$m counts \citep{kari13} in the Extended Chandra Deep Field South, AzTEC 1.1~mm combined counts of 6 blank-fields surveys \citep{scot12}, LABOCA 870~$\mu$m \citep{joha11} and SCUBA 850~$\mu$m counts around galaxy clusters \citep{knud08}. 
We also show the model predictions of \cite{shim12}, \cite{hayw13}, \cite{beth12}, and \cite{dacu13}. 
The 1.1~mm, 850~$\mu$m, and 870~$\mu$m counts are offset to an equivalent 1.3mm flux density using scaling factors of $S_{\rm 1.3 mm}/S_{\rm 1.1 mm} = 0.71$, $S_{\rm 1.3 mm}/S_{\rm 870 \mu m} = 0.38$, and $S_{\rm 1.3 mm}/S_{\rm 850 \mu m} = 0.36$. 
\label{fig:counts}}
\end{figure}

By using the 15 serendipitously-detected sources, we create cumulative number counts. 
To create number counts, we correct for the contamination of spurious sources, the incompleteness, and the survey area for each source. 
The contamination of spurious sources to each source ($C_{\rm spurious}$) is estimated from the average spurious detections at its SN (Fig.~\ref{fig:fdr}) and is subtracted from $1$ ($1 - C_{\rm spurious}$). 
Then it is divided by the completeness ($(1 - C_{\rm spurious})$/completeness), and finally corrected for the survey area by using the effective area ($(1 - C_{\rm spurious}$)/completeness/effective area). 
The upper and lower errors are calculated for each source from Poisson confidence limits of 84.13\%, which correspond to $1 \sigma$ for Gaussian statistics \citep{gehr86}, and then corrected for the incompleteness and survey area in the same manner as described above. 
The derived number counts are shown in Fig.~\ref{fig:counts} and Table~\ref{tab:counts}. 
We also plot the number counts created by using 6 $\ge$$4.0 \sigma$ sources to see the effect of the detection threshold. 
Both counts are consistent within the margin of errors.

\begin{table}
\begin{center}
\caption{Number counts at 1.3~mm derived from the $\ge$3.8$\sigma$ sources. \label{tab:counts}}
\begin{tabular}{ccc}
\tableline\tableline
$S$ & $N_{\rm source}$ & $N(>S)$ \\
(mJy) & & ($\times 10^3$ deg$^{-2}$) \\
(1) & (2) & (3) \\ 
\tableline
0.15 & 2 & $53.1^{+108}_{-38.7}$ \\
0.24 & 4 & $18.3^{+19.4}_{-69.7}$ \\
0.37 & 7 & $9.21^{+13.0}_{-46.6}$ \\
0.59 & 2 & $2.10^{+4.57}_{-1.64}$ \\
\tableline
\end{tabular}
\tablecomments{
The errors are Poisson confidence limits of 84.13\%, corresponding to $1 \sigma$ for Gaussian statistics. 
(1) Flux density; 
(2) Number of sources in the bin; 
(3) Cumulative number counts. 
}
\end{center}
\end{table}

\section{Discussion}\label{sec:discussion}

By using the ``sub-mJy sources'' detected with ALMA, we derived cumulative number counts, which probe currently the faintest end at 1~mm wavelengths. 
We compare the number counts with other observational results and model predictions.

\subsection{Comparisons with Previous Results}

We compare our number counts with previous observational results. 
In Fig.~\ref{fig:counts}, we plot number counts obtained by SCUBA 850~$\mu$m \citep{knud08}, LABOCA 870~$\mu$m \citep{joha11}, AzTEC 1.1~mm \citep{scot12}, and ALMA band 7 (870~$\mu$m) observations \citep{kari13}. 
The flux density of these counts are scaled to 1.3~mm flux density 
by using a modified black body with typical values for SMGs; spectral index of $\beta = 1.5$ and dust temperature of 35~K \citep[e.g.,][]{kova06, copp08} 
located at $z = 2.5$ \citep[e.g.,][]{chap05, yun12}.

The faintest counts are probed by using gravitationally lensing clusters at 850~$\mu$m \citep{knud08} and at 870~$\mu$m \citep{joha11}. 
The ALMA 1.3~mm counts are consistent with those counts, suggesting that both wavelengths trace the same population at faint end.

\cite{scot12} derived currently the most reliable number counts at 1.1~mm by combining data in six blank fields surveyed by AzTEC. 
The 1.3~mm counts appear to be consistent with those at 1.1~mm, except for the faintest bin of the 1.1~mm counts. 
They suggest a turnover in the Euclidean-normalized counts at $S_{\rm 1.1\ mm} \lsim 2$~mJy. 
However, our data points at fainter flux range do not show such turnover. 
We note that number counts derived from targeted observations may show an excess of source counts.

Recently, \cite{kari13} conducted ALMA 870~$\mu$m observations of 122 sources detected in the LABOCA Extended Chandra Deep Field South Submillimeter survey. 
The high-resolution observations pinpoint the source positions, and some of the LABOCA (FWHM $\sim 19''$) sources are resolved into multiple sources. 
Number counts are derived from the ALMA results without the effect of source confusion. 
Although the extrapolation of the fainter end of the 870~$\mu$m number counts appears to be lower than the 1.3~mm counts, deeper observations at the same depth in flux density as the flux ranges of 1.3~mm observations are needed to compare both results quantitatively. 

\subsection{Comparisons with Model Predictions}

We compare the 1.3~mm counts with recent model predictions with different approaches by \cite{shim12}, \cite{beth12}, and \cite{dacu13}. 
Fig.~\ref{fig:counts} illustrates that the three models agree well with the ALMA 1.3~mm number counts.

\cite{shim12} perform large cosmological hydrodynamic simulations and simulate the properties of SMGs by calculating the reprocessing of stellar light by dust grains into far-IR to millimeter wavelengths in a self-consistent manner. 
Their model reproduces number counts and clustering amplitude of SMGs. 
The ALMA number counts are consistent with the predicted number counts calculated for 238~GHz with this model. 
In the flux range of 0.1--0.6 mJy, the simulated galaxies have a stellar mass of $\sim$$5 \times 10^9$--$10^{11}$~$M_{\odot}$ and an SFR of $\sim$10--100~$M_{\odot}$~yr$^{-1}$.

The model of \cite{beth12} is based on the redshift evolution of the mass function of star-forming galaxies, specific SFR distribution at fixed stellar mass, and spectral energy distributions (SEDs) for the two star-formation modes (main-sequence and starburst). 
The main-sequence galaxies, which follow a tight correlation between stellar mass and SFR, have steady star formation, while starburst galaxies, which are outliers to the correlation, have high specific SFRs possibly driven by mergers \citep[e.g.,][]{noes07}. 
Their model reproduces IR to radio number counts and counts per redshift slice of {\sl Herschel} observations. 
At fainter flux density ($S_{\rm 1.1mm} \lsim 1$~mJy) main-sequence galaxies dominate the number counts by an order of magnitude, suggesting that the ALMA sub-mJy sources are more like normal star-forming galaxies than classical SMGs with intense star-forming activity.

\cite{dacu13} create self-consistent models of the observed optical/NIR SEDs of galaxies detected in the Hubble Ultra Deep Field. 
They combine the attenuated stellar spectra with a library of IR emission models, which are consistent with the observed optical/NIR emission in terms of energy balance, and estimate the continuum flux at millimeter and submillimeter. 
The predicted number counts agree with the previous observed number counts and extragalactic background light. 
They simulate deep ALMA observations with a continuum sensitivity of $1 \sigma_{\rm continuum} = 5.1 \times 10^{-3}$ mJy. 
In their simulation, the redshift distribution of detected galaxies in the ALMA band 6 is relatively flat up to $z = 5$. 
The detected galaxies typically have a stellar mass of $10^9$--$10^{11}$~$M_{\odot}$ and an SFR of a few $\times$ (0.1--10)~$M_{\odot}$~yr$^{-1}$, which are more massive, more star-forming, and more dusty galaxies among their optical/NIR-detected galaxy sample. 
The agreement between the model and our ALMA number counts implies that the sub-mJy sources are massive and high-SFR end of optical/NIR-detected galaxies.

\subsection{Contribution to the Extragalactic Background Light}

We calculate the fraction of the EBL revolved into discrete sources in the ALMA 1.3~mm observations. 
The total flux density of the $\ge$3.8$\sigma$ sources, which is applied all the corrections conducted in the derivation of the number counts (Section~{\ref{sec:source}}), is $\sim$13~Jy~deg$^{-2}$. 
The EBL at 1.3~mm based on the measurements by the {\sl Cosmic Background Explorer} satellite is 14--18 Jy~deg$^{-2}$ \citep{puge96, fixs98}. 
Therefore, we have resolved $\sim$80\% of the EBL into discrete sources, which is significantly higher compared to results in the previous deep surveys (10\%--20\%) at 1~mm wavelengths \citep[e.g.,][]{grev04, scot10, hats11}, thanks to the high sensitivity and high angular resolution of ALMA. 
The index of a power-law fit to the number counts is steeper than $-1$, indicating that the contribution of fainter sources is significant.

\section{Summary}\label{sec:summary}

We conducted deep 1.3~mm observations with ALMA toward 20 star-forming galaxies at $z \sim 1.4$ in the SXDS field. 
We serendipitously detected 15 sources ($\ge$3.8$\sigma$) other than the targeted star-forming galaxies. 
The flux densities of the ALMA sources are $S_{\rm 1.3\ mm} = 0.15$--0.61 mJy, which are an order of magnitude fainter than those of SMGs detected in previous surveys. 
By using these `sub-mJy sources', we created number counts, which probe the faintest flux range at millimeter wavelengths. 
The number counts are consistent with flux-scaled 850~$\mu$m and 870~$\mu$m counts obtained with gravitational lensing clusters, which trace the faint end of number counts at submillimeter wavelengths. 
The number counts agree well with model predictions, which suggest that the sub-mJy sources are moderately star-forming galaxies with SFR $\lsim$ 100~$M_{\odot}$~yr$^{-1}$ rather than classical SMGs with intense star formation. 
We find that our ALMA observations resolves $\sim$80\% of the EBL at 1.3~mm.

\acknowledgments

We would like to acknowledge Kazuya Saigo and staffs at the ALMA Regional Center for their help in data reduction. 
We are grateful to Ikko Shimizu for providing their model counts and for useful comments, and Kotaro Kohno, Kouichiro Nakanishi, and Daisuke Iono for useful discussions. 
We thank the referee for helpful comments and suggestions. 
BH is supported by Research Fellowship for Young Scientists from the Japan Society of the Promotion of Science (JSPS). 
KO is supported by the Grant-in-Aid for Scientific Research (C)(24540230) from JSPS. 
MA is supported by the Grant-in-Aid for Scientific Research (B)(21340042) from JSPS.
This paper makes use of the following ALMA data: ADS/JAO.ALMA\#2011.0.00648.S. 
ALMA is a partnership of ESO (representing its member states), NSF (USA) and NINS (Japan), together with NRC (Canada) and NSC and ASIAA (Taiwan), in cooperation with the Republic of Chile. 
The Joint ALMA Observatory is operated by ESO, AUI/NRAO and NAOJ.

{\it Facilities:} \facility{ALMA}.


\end{document}